\documentstyle[12pt]{article}

\begin{document}
\baselineskip=20pt

\begin{titlepage}

\begin{flushright}
\begin{minipage}[t]{2.4cm}
\begin{flushleft}
DPNU-93-01 \\
KUNS-1178 \\
March, 1993
\end{flushleft}
\end{minipage}
\end{flushright}

\vspace{0.7cm}

\begin{center}
\Large\bf
Proving the Low Energy Theorem \\
of Hidden Local Symmetry
\end{center}

\vfill

\begin{center}
Masayasu {\sc Harada}%
\footnote{Address after April 1, 1993 :
Department of Physics, Kyoto University,
Kyoto 606-01, Japan}\\
{\it
Department of Physics,
Nagoya University,\\
Nagoya 464-01, Japan }\\
\ \\
Taichiro {\sc Kugo}\\
{\it
Department of Physics,
Kyoto University,\\
Kyoto 606-01, Japan}\\
\ \\
Koichi {\sc Yamawaki}\\
{\it
Department of Physics,
Nagoya University,\\
Nagoya 464-01, Japan}\\
\end{center}

\vspace{0.5cm}
\vfill

{\small
\begin{center}
{\bf Abstract\vspace{-.5em}\vspace{0pt}}
\end{center}
\begin{list}{}{%
\setlength{\listparindent}{1.5em}
\setlength{\itemindent}{\listparindent}
\setlength{\leftmargin}{5em}
\setlength{\rightmargin}{\leftmargin}
\setlength{\parsep}{0pt plus 1pt}
}\item[]
Based on the Ward-Takahashi identity for the BRS symmetry,
we prove to all orders of the loop expansion the low energy theorem
of hidden local symmetry for the vector mesons (KSRF (I) relation)
in the $U(N)_{\rm L}$ $\times$ $U(N)_{\rm R}$ /
$U(N)_{\rm V}$ nonlinear
chiral Lagrangian.
\end{list}
}
\vspace{1cm}
\end{titlepage}

Hidden local symmetry (HLS) is a natural framework for describing
the vector mesons in a manner consistent with the chiral
symmetry of QCD\cite{BandoKugoYamawaki}.
The HLS Lagrangian yields at tree level
a successful phenomenology for the pions and the
$\rho$ mesons.
By choosing a parameter $a=2$ in this HLS Lagrangian,
we have\cite{BandoKugoUeharaYamawakiYanagida}:
1) Universality of the $\rho$ meson coupling\cite{Sakurai}
$
g_{\rho \pi \pi} = g
$
($g$: HLS gauge coupling);
2) KSRF relation\cite{KSRF} (version II)
$
m_{\rho}^2 = 2 f_{\pi}^2 g_{\rho \pi \pi}^2
$;
3) $\rho$ meson dominance of the electromagnetic
form factor of the pion
$
g_{\gamma \pi \pi} = 0
$\cite{Sakurai}.

Most remarkably, we further obtain an $a$-independent
relation\cite{BandoKugoYamawakiNP}
$
g_{\rho} / g_{\rho \pi \pi} = 2 f_{\pi}^2
$,
with $g_{\rho}$ being the strength of $\rho$-$\gamma$ mixing.
This is actually another version of the celebrated
KSRF relation (KSRF I) and is a decisive test of the HLS
in the hadron physics.
Since it follows from the symmetry structure
alone, it was conjectured\cite{BandoKugoYamawakiNP}
to be a low energy theorem valid at zero momenta for
{\it any} Lagrangian possessing the HLS and was further proved
\cite{BandoKugoYamawakiPTP} at tree level.
If it is indeed a genuine low energy theorem
surviving the loop corrections,
analogues of such a relation would also be useful for
strongly coupled Higgs models or
models of the dynamical electroweak symmetry breaking.
Actually this HLS Lagrangian can be straightforwardly applied to
such models\cite{Casaletal}.

Quite recently, it was shown in Landau gauge\cite{HaradaYamawaki}
that the above ``low energy theorem" as well as the $a=2$ tree-level
results at zero momenta is not altered by the one-loop corrections,
thus strongly suggesting that it may indeed be a true low energy
theorem\cite{foot:offshell}.
These results are actually the relations coming from
$O(p^2)$ operators or dimension-2 operators (counting only
dimensions of the gauge fields and the derivatives).

In this paper we shall prove that the above
``low energy theorem'' of HLS actually holds at any loop order,
based on the Ward-Takahashi (WT) identity for the
Becchi-Rouet-Stora (BRS) symmetry.
By restricting ourselves to the dimension-2 operators mentioned above,
we can prove it by mathematical induction in quite the same way as the
renormalizability proof for gauge theories\cite{BRS}
and two-dimensional
nonlinear sigma models\cite{BlasiCollina}.
In the following we assume that
there exists a gauge invariant regularization
(for example, the dimensional regularization).

Let us start with the
$G_{\rm global}\times H_{\rm local}$
``linear model",
with $G=U(N)_{\rm L} \times U(N)_{\rm R}$
and $H=U(N)_{\rm V}$.
The Lagrangian is given by\cite{BandoKugoUeharaYamawakiYanagida}
\begin{eqnarray}
  {\cal L}
&=&
  {\cal L}_{\rm A} + a {\cal L}_{\rm V} +
  {\cal L}_{\rm kin} (V_\mu) ,
\nonumber\\
&{}&
  {\cal L}_{{\rm A}\atop{\rm V}} =
  -{f_\pi^2 \over 4}
  \mbox{\rm tr}
  \left[{
    \left({
      D_\mu \xi_{\rm L} \cdot \xi_{\rm L}^{\dag}
      \mp D_\mu \xi_{\rm R} \cdot \xi_{\rm R}^{\dag}
    }\right)^2
  }\right]
  ,
\label{StartLag}
\end{eqnarray}
where $a$ is a constant and
${\cal L}_{\rm kin}(V_\mu)$ denotes the
kinetic term
of the hidden gauge boson
(vector meson).
Here $\xi_{\rm L}(x)$ and $\xi_{\rm R}(x)$
are two $U(N)$-matrix valued variables,
which transform as
$  \xi_{\rm L,R}(x) \rightarrow
  \xi'_{\rm L,R}(x) = h(x) \xi_{\rm L,R}(x) g_{\rm L,R}^{\dag}
$,
where
$ h(x) \in H_{\rm local}$
and
$ g_{\rm L,R} \in G_{\rm global}$.
These variables are parameterized as
$
 \xi_{\rm L,R}(x) \equiv \exp[i \phi_{\rm L,R}^a (x) T_a ],
$
where $T_a$ denotes the $U(N)$ generator.
The covariant derivatives of $\xi_{\rm L,R}(x)$ are defined by
\begin{equation}
  D_\mu \xi_{\rm L}
  \equiv
  \partial_\mu \xi_{\rm L} - i  V_\mu \xi_{\rm L} +
   i \xi_{\rm L} {\cal V}_{{\rm L}\mu}  ,
\end{equation}
and similar one
with replacement
${\rm L} \leftrightarrow {\rm R}$,
${\cal V}_{{\rm L}\mu} \leftrightarrow {\cal V}_{{\rm R}\mu}$,
where $V_\mu$ ($\equiv V_\mu^a T_a$)
is the hidden gauge boson field
and ${\cal V}_{{\rm L}\mu}$ and
${\cal V}_{{\rm R}\mu}$ denote the external gauge fields
gauging the
$ G_{\rm global}$ symmetry.

Now we consider the
loop effects of the model.
Let us take a covariant gauge condition
for the HLS,
and introduce the corresponding gauge-fixing
and Faddeev-Popov (FP) terms:
\begin{equation}
  {\cal L}_{GF} + {\cal L}_{FP}
  =
  B^a \partial^\mu V_\mu^a
  + {1\over2} \alpha B^a B^a
  + i \bar{ C}^a \partial^\mu D_\mu  C^a ,
\label{GFFPlagrangian}
\end{equation}
where $B^a$ is the Nakanishi-Lautrap (NL) field and
$ C^a$ ($\bar{ C}^a$) the FP ghost (anti-ghost) field.
In this paper we do not consider
the radiative corrections due to the external gauge fields
${\cal V}_\mu^i \equiv ({\cal V}_{{\rm L}\mu}^a ,
{\cal V}_{{\rm R}\mu}^a)$,
so that we need not introduce the gauge-fixing terms
for ${\cal V}_\mu$.
Then, the corresponding ghost fields
${\cal C}^i\equiv({\cal C}_{\rm L}^a,{\cal C}_{\rm R}^a)$
are non-propagating.

The infinitesimal form of the
$ G_{\rm global} \times  H_{\rm local}$ transformation
of the Nambu-Goldstone (NG) field $\phi^i\equiv(\phi_L^a,\phi_R^a)$
is given by
$
  \delta \phi^i
  =
  \theta^a W_a^i(\phi)
  + \vartheta^j {\cal W}_j^i (\phi)
$
$
  \left({
    \equiv \theta^A \hbox{\boldmath $W$}_A^i(\phi)
  }\right)
$,
where $A$ denotes a set $(a,i)$ of labels of $ H_{\rm local}$ and
$ G_{\rm global}$.
Accordingly,
the BRS transformation of the NG fields $\phi^i$,
the gauge fields
$\hbox{\boldmath $V$}_\mu^A \equiv (V_\mu^a,{\cal V}_\mu^i)$
and the FP ghost fields
$\hbox{\boldmath $C$}^A\equiv( C^a,{\cal C}^i)$
are respectively given by
\begin{eqnarray}
  \delta_{\rm B} \phi^i
&=&
  \hbox{\boldmath $C$}^A \hat{\hbox{\boldmath $W$}}_{\!\!A} \phi^i
  \quad
  ( \hat{\hbox{\boldmath $W$}}_{\!\!A} \equiv
  \hbox{\boldmath $W$}_{\!A}^i (\phi)
  {\partial \over \partial \phi^i} ) ,
\nonumber\\
  \delta_{\rm B} \hbox{\boldmath $V$}_\mu^A
&=&
  \partial_\mu \hbox{\boldmath $C$}^A +
  \hbox{\boldmath $V$}_\mu^B \hbox{\boldmath $C$}^C {f_{BC}}^A ,
\nonumber\\
  \delta_{\rm B} \hbox{\boldmath $C$}^A
&=&
  - {1\over2} \hbox{\boldmath $C$}^B
  \hbox{\boldmath $C$}^C {f_{BC}}^A .
\end{eqnarray}

We define the dimension of the fields as
$\dim[\phi_{L,R}^a]=0$,
$\dim[\hbox{\boldmath $V$}_\mu^A]=1$.
It is also convenient to assign the following dimensions to
the FP-ghosts:
$\dim[\hbox{\boldmath $C$}^A]=0$,
$\dim[\bar{ C}^a]=2$.
Then the BRS transformation does not change the dimension.
According to the above dimension counting,
we may divide
the Lagrangian Eq.(\ref{StartLag}) plus
Eq.(\ref{GFFPlagrangian})
into two parts:\
1)\  dimension-2 part ${\cal L}_{\rm A} + a {\cal L}_{\rm V}$ and
2)\  dimension-4 part
${\cal L}_{\rm kin}(V_\mu) + {\cal L}_{\rm GF} + {\cal L}_{\rm FP}$.
(We count the dimension of the fields
and derivatives only, hereafter.)

Now, we consider the quantum correction
to this system at any loop order, and
prove the following proposition.

{\bf Proposition} :\
{\it As far as the dimension-2 operators are concerned,
all the quantum corrections,
including the finite parts as well as the divergent parts,
can be absorbed into the original dimension-2 Lagrangian
${\cal L}_{\rm A} + a {\cal L}_{\rm V}$
by a suitable redefinition (renormalization) of the parameters
$a$, $f_\pi^2$, and the fields $\phi^i$,
$V_\mu^a$.}

\noindent
This implies that the tree-level dimension-2-Lagrangian,
with the parameters and fields
substituted by the ``renormalized" ones,
already
describes the exact action at any loop order,
and therefore that all the ``low energy theorems"
derived from it
receive no quantum corrections at all.

We prove the above proposition
in the same way as the renormalizability proof
for gauge theories\cite{BRS} and
two dimensional nonlinear sigma models\cite{BlasiCollina}.
We can write down the WT identity
for the effective action $\Gamma$.
The NL fields $B^a$ and
the FP anti-ghost fields $\bar{ C}^a$
can be eliminated from $\Gamma$
by using their equations of motion as usual.
Then the tree level action $S=\Gamma_{\rm tree}$ reads
\begin{eqnarray}
&{}&
  S [ \Phi, {\bf K}; \hbox{\boldmath $a$} ]
  =
  S_2 [\phi,\hbox{\boldmath $V$}] + S_4[\Phi,{\bf K}] ,
\nonumber\\
&{}& \qquad
  S_2 [\phi,\hbox{\boldmath $V$}] = \int d^4x
  \left({
    a_{\perp} {\cal L}_{\rm A} (\phi,\hbox{\boldmath $V$}) +
     a_{\parallel} {\cal L}_{\rm V} (\phi,\hbox{\boldmath $V$})
  }\right) ,
\nonumber\\
&{}& \qquad
  S_4 [\Phi,{\bf K}] = \int d^4x
  \left({
    {\cal L}_{\rm kin}(V_\mu)
     + {\bf K} \cdot \delta_{\rm B} \Phi
  }\right) ,
\label{SfourAction}
\end{eqnarray}
where
$\Phi \equiv (\phi^i,\hbox{\boldmath $V$}_\mu^A,
\hbox{\boldmath $C$}^A)$ are
the field variables and
${\bf K}\equiv(K_i,\hbox{\boldmath $K$}_A^\mu,
\hbox{\boldmath $L$}_A)$
( $ \hbox{\boldmath $K$}_A^\mu \equiv (K_a^\mu,{\cal K}_i^\mu)$,
$\hbox{\boldmath $L$}_A \equiv (L_a,{\cal L}_i)$ )
denote the BRS source fields.
We have rewritten $a$ and $f_\pi^2$
as $a f_\pi^2 \rightarrow  a_{\parallel} f_\pi^2$ and
$f_\pi^2\rightarrow a_{\perp} f_\pi^2$,
so that the renormalization of $a$ and $f_\pi^2$
corresponds to that of
$\hbox{\boldmath $a$}\equiv( a_{\parallel},a_{\perp})$.
According to the dimension assignment of the fields,
the dimension of the above BRS source fields ${\bf K}$ is given by
$\dim[K_i]=\dim[\hbox{\boldmath $L$}_A]=4$
and $\dim[\hbox{\boldmath $K$}_A^\mu]=3$.

The WT identity for the effective action
$\Gamma$ is given by
\begin{equation}
  \Gamma\ast\Gamma=0 ,
\end{equation}
where the $\ast$ operation is defined by
\begin{equation}
  F \ast G
  =
  {(-)}^{\Phi}
  {\overleftarrow{\delta} F \over \delta \Phi }
  {\delta G \over \delta {\bf K}}
  - {(-)}^{\Phi}
  {\overleftarrow{\delta} F \over \delta {\bf K}}
  {\delta G \over \delta \Phi}
\end{equation}
for arbitrary functionals $F[\Phi,{\bf K}]$
and $G[\Phi,{\bf K}]$.
(Here the symbols $\delta$ and $\overleftarrow{\delta}$
denote the derivatives from the left and right, respectively,
and ${(-)}^\Phi$ denotes $+1$ or $-1$
when $\Phi$ is bosonic or fermionic, respectively.)

The effective action is calculated in the loop expansion:
$\Gamma = S + \hbar \Gamma^{(1)} + \hbar^2 \Gamma^{(2)} + \cdots $.
The $\hbar^n$ term $\Gamma^{(n)}$
contains contributions not only from the genuine $n$-loop diagrams
but also from the lower loop diagrams
including the counter terms.
We can expand the $n$-th term
$\Gamma^{(n)}$ according to the dimension:
\begin{equation}
  \Gamma^{(n)} =
  \Gamma_0^{(n)} [\phi] + \Gamma_2^{(n)}[\phi,\hbox{\boldmath $V$}]
  + \Gamma_4^{(n)}[\Phi,{\bf K}]
  + \cdots .
\label{dimexp}
\end{equation}
Here again we are counting the dimension only of the fields
and derivatives.
The first dimension-0 term $\Gamma_0^{(n)}$ can contain only
the dimensionless field $\phi^i$ without derivatives.
The two dimensions of the second term $\Gamma_2^{(n)}$ is
supplied by derivative and/or the gauge field
$\hbox{\boldmath $V$}_\mu^A$.
The BRS source field ${\bf K}$ carries dimension 4 or 3,
and hence it can appear only in $\Gamma_4^{(n)}$ and beyond:
the dimension-4 term $\Gamma_4^{(n)}$
is at most linear in ${\bf K}$,
while the dimension-6 term $\Gamma_6^{(n)}$ can
contain a quadratic term in $K_a^\mu$,
the BRS source of the hidden gauge boson $V_\mu^a$.
To calculate $\Gamma^{(n)}$,
we need to use the ``bare" action,
$
  {(S_0)}_n = S
  \left[{
    {(\Phi_0)}_n , {({\bf K}_0)}_n ; {(\hbox{\boldmath $a$}_0)}_n
  }\right]
$,
where the $n$-th loop order ``bare" fields
${(\Phi_0)}_n$, ${({\bf K}_0)}_n$
and parameters ${(\hbox{\boldmath $a$}_0)}_n$ are given by
\begin{eqnarray}
 {(\Phi_0)}_n
&=&
  \Phi + \hbar \delta \Phi^{(1)} +
  \cdots + \hbar^n \delta \Phi^{(n)} ,
\nonumber\\
  {({\bf K}_0)}_n
&=&
  {\bf K} + \hbar \delta {\bf K}^{(1)}
  + \cdots + \hbar^n \delta {\bf K}^{(n)} ,
\nonumber\\
  {(\hbox{\boldmath $a$}_0)}_n
&=&
  \hbox{\boldmath $a$} + \hbar \delta \hbox{\boldmath $a$}^{(1)}
 + \cdots + \hbar^n \delta \hbox{\boldmath $a$}^{(n)} .
\end{eqnarray}

Let us now prove the following by mathematical induction
with respect
to the loop expansion parameter $n$:

\begin{enumerate}
\renewcommand{\labelenumi}{\arabic{enumi})}

\item $\Gamma_0^{(n)} (\phi) = 0$.

\item
By choosing suitably the $n$-th order counter terms
$\delta\Phi^{(n)}$, $\delta{\bf K}^{(n)}$ and
$\delta\hbox{\boldmath $a$}^{(n)}$,
$\Gamma_2^{(n)}[\phi,A]$ and the ${\bf K}$-linear terms
in $\Gamma_4^{(n)}[\Phi,{\bf K}]$ can be made vanish;\
$
 \Gamma^{(n)}_2 [\phi,\hbox{\boldmath $V$}] \! = \!
 \Gamma^{(n)}_4 [\Phi,{\bf K}] %
 \vert_{{\bf K}\hbox{-}{\rm linear}} \! = 0
$.

\item The field reparameterization (renormalization)
$(\Phi,{\bf K})\rightarrow
\left({ {(\Phi_0)}_n,{({\bf K}_0)}_n }\right)$ is a ``canonical"
transformation which leaves the $\ast$ operation invariant.

\end{enumerate}

Suppose that the above statements are satisfied
for the $(n-1)$-th loop order effective action $\Gamma^{(n-1)}$.
We calculate, for the moment,
the $n$-th loop effective action $\Gamma^{(n)}$
using the $(n-1)$-th loop level ``bare" action
${(S_0)}_{n-1}$,
i.e., without $n$-th loop counter terms.
We expand
the $\hbar^n$ terms
in the WT identity
$ S \ast \Gamma^{(n)} = - {1\over2}
\sum_{l=1}^{n-1} \Gamma^{(l)} \ast \Gamma^{(n-l)}$
according to the dimensions
like in Eq.(\ref{dimexp}).
Then using the above induction assumption,
we find:
$  S_4 \ast \Gamma_0^{(n)} + S_2 \ast \Gamma_2^{(n)} = 0$
(dim 0),
$  S_4 \ast \Gamma_2^{(n)} + S_2 \ast \Gamma_4^{(n)} = 0$
(dim 2) and
$  S_4 \ast \Gamma_4^{(n)} + S_2 \ast \Gamma_6^{(n)} = 0$
(dim 4).
These three renormalization equations give enough
information for determining possible forms of
$\Gamma_0^{(n)}$, $\Gamma_2^{(n)}$ and
$\Gamma_4^{(n)}\vert_{{\bf K}\hbox{-}{\rm linear}}$
(the ${\bf K}$-linear term in $\Gamma_4^{(n)}$)
which we are interested in.

First,
the dimension-0 part
of the renormalization equation reads
$\delta_{\rm B}\Gamma_0^{(n)}=0$.
Since there are no invariants containing no derivatives,
we can immediately conclude $\Gamma_0^{(n)}=0$,
and hence our statement 1) follows.

Next, we solve the dimension-2 and the dimension-4 parts of
the above renormalization equations.
It is convenient to define the BRS-like transformation
$\delta'_\Gamma$ on the fields $\Phi$ by
$ \delta'_\Gamma \equiv
\left( \delta \Gamma^{(n)}_4 / \delta {\bf K} \right)
{\delta \over \delta \Phi}$.
Then these equations read
\begin{eqnarray}
&{}&
  \delta_{\rm B} \Gamma_2^{(n)} + \delta'_\Gamma S_2 = 0 ,
\label{DimtwoWTB} \\
&{}&
  \delta_{\rm B} \Gamma_4^{(n)} + \delta'_\Gamma S_4
  + {\delta \Gamma_6^{(n)} \over \delta {\bf K}}
    {\delta S_2 \over \delta \Phi} = 0 .
\label{DimfourWTB}
\end{eqnarray}

A tedious but straightforward analysis\cite{foot:det}
of the ${\bf K}$-linear term in Eq.(\ref{DimfourWTB})
determines
the general form of
the $\Gamma_4^{(n)}\vert_{{\bf K}\hbox{-}{\rm linear}}$
and $\Gamma_6^{(n)}\vert_{{\bf K}\hbox{-}{\rm quadratic}}$ terms:
the solution for $\Gamma_4^{(n)}\vert_{{\bf K}\hbox{-}{\rm linear}}$
or equivalently $\delta'_\Gamma$
is given by
\begin{eqnarray}
&{}&
  \delta'_\Gamma  C^a = \beta \delta_{\rm B}  C^a ,
\\
&{}&
  \delta'_\Gamma \phi^i
  =
  \left\{{
     C^a
    \left({
      [ \hat{W}_a,\hat{F} ] + \beta \hat{W}_a
    }\right)
    + {\cal C}^j [\hat{\cal W}_j,\hat{F}]
  }\right\}\phi^i ,
\\
&{}&
  \delta'_\Gamma V_\mu^a
  =
  \alpha \partial_\mu  C^a
  + \beta \delta_{\rm B} V_\mu^a
  + \gamma \delta_{\rm B}
  \left({ V_\mu^a - \tilde{\cal V}_\mu^a }\right) ,
\end{eqnarray}
where $\alpha$, $\beta$ and $\gamma$ are constants,
$
\tilde{\cal V}_\mu \equiv
  \xi_{\rm L} {\cal V}_{{\rm L}\mu} \xi_{\rm L}^{\dag}
  -i \partial_\mu\xi_{\rm L} \cdot \xi_{\rm L}^{\dag}
+ ({\rm L} \leftrightarrow {\rm R})
$
and
$\hat{F} \equiv F^i(\phi) \partial/\partial\phi^i$,
with $F^i(\phi)$ being a certain dimension-0 function.
Note that $\delta'_\Gamma{\cal V}_\mu^i=\delta'_\Gamma{\cal C}^i=0$,
since the external $ G_{\rm global}$-gauge
fields ${\cal V}_\mu^i$ and
their ghosts ${\cal C}^i$ are not quantized
and hence their BRS source fields
${\cal K}_i^\mu$ and ${\cal L}_i$ appear only in the tree action.

Using $\delta'_\Gamma$ thus obtained,
we next solve the above WT identity Eq.(\ref{DimtwoWTB})
and easily find
\begin{equation}
  \Gamma_2^{(n)}
  =
  A_{2{\rm GI}} [\phi,\hbox{\boldmath $V$}]-
  \left({
    \hat{F} S_2 + \alpha V_\mu^a
    {\delta \over \delta V_\mu^a} S_2
  }\right),
\end{equation}
where $A_{2{\rm GI}}$ is a dimension-2
gauge-invariant function of $\phi^i$ and
$\hbox{\boldmath $V$}_\mu^A$.

The solutions
are combined into a simple form
\begin{equation}
  \Gamma_2^{(n)} +
  \left. \Gamma_4^{(n)} \right\vert_{{\bf K}\hbox{-}{\rm linear}}
  =
  A_{2{\rm GI}} [\phi,\hbox{\boldmath $V$}] - S \ast Y
\label{Solution}
\end{equation}
up to irrelevant terms
(dimension-6 or ${\bf K}$-independent dimension-4 terms),
where
the functional $Y$ is given by
\begin{eqnarray}
  Y
&=&
  \int d^4x
  \bigl[
    K_i F^{i} (\phi)
    + \alpha K_a^\mu V_\mu^a
\nonumber\\
&{}& ~~~~~~~~~~
    + \beta L_a  C^a
    + \gamma f_{abc} K_a^\mu K_{b\mu}  C^c
  \bigr] .
\label{SolY}
\end{eqnarray}

Now, we prove our statements 2) and 3) in the above.
We have calculated the above effective action $\Gamma^{(n)}$
{\it without} using $n$-th loop level counter terms
$\delta\Phi^{(n)}$, $\delta{\bf K}^{(n)}$ and
$\delta\hbox{\boldmath $a$}^{(n)}$.
If we include those,
we have the additional contribution given by
\begin{equation}
  \Delta \Gamma^{(n)}
  =
  \delta\Phi^{(n)} {\delta S \over \delta \Phi}
  + \delta{\bf K}^{(n)} {\delta S \over \delta {\bf K}}
  + \delta\hbox{\boldmath $a$}^{(n)}
    {\partial S \over \partial \hbox{\boldmath $a$}} ,
\label{counter}
\end{equation}
where $S[\Phi,{\bf K};\hbox{\boldmath $a$}]$
is the tree-level action.
So the true $n$-th loop level effective action is given by
$\Gamma^{(n)}+\Delta\Gamma^{(n)}
\equiv\Gamma^{(n)}_{{\rm total}}$.
The tree-level action $S_2$
is the most general gauge-invariant dimension-2 term,
so that
$A_{2{\rm GI}}[\phi,\hbox{\boldmath $V$}]$ term
in Eq.(\ref{Solution})
can be canceled
by suitably chosen counter terms,
$\delta\hbox{\boldmath $a$}^{(n)} {\partial S \over \partial
\hbox{\boldmath $a$}}$.
The
second term
$-S\ast Y$ term in Eq.(\ref{Solution})
just represents
a ``canonical transformation'' of $S$
generated by $-Y$.
Therefore we choose
the $n$-th
order field counter terms $\delta\Phi^{(n)}$
and $\delta{\bf K}^{(n)}$ to be equal to the canonical
transformations of $\Phi$ and ${\bf K}$ generated by $+Y$;
$\delta\Phi^{(n)} = \Phi \ast Y$,
$\delta{\bf K}^{(n)} = {\bf K} \ast Y$.
Then
the first and the second terms in Eq.(\ref{counter})
just give $S \ast Y$
and precisely cancel
the second term in Eq.(\ref{Solution}).
Thus we have completed the proof of our statements 2) and 3).

Some comments are in order:

\noindent
1.~
Our conclusion in this paper remains unaltered
even if the action $S$ contains other dimension-4 or higher terms,
as far as they respect the symmetry.
This is because we needed just
$\left({S \ast \Gamma}\right)_2$
and $\left({S \ast \Gamma}\right)_4 %
\vert_{{\bf K}\hbox{-}{\rm linear}}$
parts in the WT identity
to which only
$S_2$ and ${\bf K}\hbox{-}{\rm linear}$ part
of $S_4$ can contribute.

\noindent
2.~
When we regard this HLS model
as a low energy effective theory of QCD,
we must
take account of
the anomaly
and the corresponding
Wess-Zumino-Witten term $\Gamma_{\rm WZW}$.
The WT identity
now reads
$\Gamma\ast\Gamma=({\rm anomaly})$.
However,
the RHS
is saturated already at the tree level
in this effective Lagrangian
and
so the WT identity
at loop levels,
which we need,
remains the same as before.
The WZW term $\Gamma_{\rm WZW}$ or
any other intrinsic-parity-odd terms\cite{FKTUY} in $S$
are of dimension-4 or higher and hence
do not change our conclusion as explained above.

\noindent
3.~
We have shown in the covariant gauges
that our tree-level dimension-2 action
$\int d^4 x ({\cal L}_{\rm A} + a {\cal L}_{\rm V} ) $,
if written in terms of renormalized parameters and fields,
already gives the exact action $\Gamma_2$
including all the loop effects. This form of the effective action
(in particular the ${\cal L}_{\rm V}$ part) implies that
the previously derived
relation\cite{BandoKugoYamawakiNP,BandoKugoYamawakiPTP}
\begin{equation}
{g_V (p^2) \over
g_{V\pi\pi}(p^2, p_{\pi_1}^2\!\!=\! p_{\pi_2}^2\!\!=\!0)}
\bigg\vert_{p^2=0} = 2f_\pi^2
\label{eqLET}
\end{equation}
{\it is} actually an exact low energy theorem valid at any loop
order. Of course, this theorem concerns off-shell quantities at
$p^2=0$ of the vector field momentum $p$,
and hence is not physical as it stands.
However, suppose that the vector mass
$m_V^2=ag^2f_\pi^2$ is sufficiently small compared
with the characteristic energy scale $\Lambda^2$ of the system,
which is customarily taken as $\Lambda^2\sim 16\pi^2 f_\pi^2$.
Then we expect that the on-shell value of
$g_V/g_{V\pi\pi}$ at $p^2=m_V^2$ can deviate
from the LHS of Eq.(\ref{eqLET}) only by a quantity of order
$m_V^2/\Lambda^2\sim ag^2/16\pi^2$, since the contributions of the
dimension-4 or higher terms in the effective
action $\Gamma$ (again representing all the loop effects) are
suppressed by a factor of $p^2/\Lambda^2$ at least.
Therefore as far as the vector mass is light,
namely, when either $a$ or $g^2/16\pi^2$ is small,
our theorem is truly a physical one\cite{foot:veclimit}.
In the actual world of QCD, the $\rho$ meson mass
is not so light ($ag^2/16\pi^2\sim 1/2$) so that the situation
becomes a bit obscure. Nevertheless,
the fact that the KSRF (I) relation
$g_{\rho}/g_{\rho\pi\pi}=2f_\pi^2$
holds on the $\rho$ mass shell with good accuracy
strongly suggests that the $\rho$ meson is
the hidden gauge field
and {\it the KSRF (I) relation is a physical manifestation of our
low energy theorem}.

\noindent
4.~
In this connection we should comment on the gauge choice.
In the covariant gauges which we adopted here,
the $ G_{\rm global}$ and $ H_{\rm local}$ BRS symmetries
are separately preserved.
Accordingly,
the $V_\mu$ field is multiplicatively
renormalized (recall that
$\delta V_\mu^{(n)} = V_\mu \ast Y = \alpha V_\mu$),
and the above (off-shell) low energy theorem Eq.(\ref{eqLET}) holds.
However, if we adopt $R_\xi$-gauges
(other than Landau gauge),
these properties are violated;
for instance, $\phi\partial_\mu \phi$ or the external gauge field
${\cal V}_\mu$ gets mixed with our $V_\mu$
through the renormalization,
and our off-shell low energy theorem Eq.(\ref{eqLET}) is violated.
This implies that the $V_\mu$ field in the $R_\xi$ gauge
generally does not give a smooth off-shell extrapolation;
indeed, in $R_\xi$ gauge with gauge parameter
$\alpha \equiv 1/\xi$, the correction
to $g_{\rho}/g_{\rho\pi\pi}$ by
the extrapolation from $p^2=m_\rho^2$ to $p^2=0$ is seen
to have a part proportional to $\alpha g^2/16\pi^2$, which diverges
when $\alpha$ becomes very large.
Thus, in particular, the unitary gauge\cite{foot:unitary},
which corresponds to
$\alpha \rightarrow \infty$,
gives an ill-defined off-shell field.

\noindent
5.~
Our argument is free from infrared divergences
at least in Landau gauge.
This can be seen as follows.
In this gauge
the propagators of the NG bosons, the hidden gauge bosons
and the FP ghosts
(after rescaling
the FP anti-ghost
$\bar{C}$ into $f_\pi^2\bar{C}$)
are all proportional to $1/f_\pi^2$ in the infrared region.
Therefore, a general $L$-loop diagram,
which includes $V_4$ dimension-4 vertices
and $K$ BRS source vertices,
yields an amplitude proportional to
$\left({1/f_\pi^2}\right)^{(L-1+V_4+K)}$\cite{foot:ChPT}.
Thus, from dimensional consideration
we see that there is no infrared contribution to
$\Gamma_0^{(n)}[\phi]$,
$\Gamma_2^{(n)}[\phi,\hbox{\boldmath $V$}]$ and
$\Gamma_4^{(n)}[\Phi,{\bf K}]\vert_{{\bf K}\hbox{-}{\rm linear}}$.
In other covariant gauges,
there appears a dipole ghost
in the vector propagator,
which is to be defined by a suitable regularization.

We would like to thank Volodya Miransky and
Masaharu Tanabashi for stimulating discussions.
T.~K. is supported in part by the Grant-in-Aid for
 Scientific Research (\#04640292) from the Ministry of Education,
Science and Culture.
K.~Y. is supported in part by the Takeda Science Foundation
and the Ishida Foundation,
and also by the International Collaboration Program
of the Japan Society for Promotion of Science.


\end{document}